\documentclass[preprint,showpacs,showkeys,amsmath,amssymb,aps,doublespace]{revtex4-1}
\usepackage[dvips]{graphicx}
\usepackage{setspace}
\usepackage{amsmath}
\usepackage{amsfonts}
\usepackage{amssymb,color}
\usepackage{color}
\usepackage{ifpdf}
\usepackage[latin1]{inputenc}

\begin{document}

\title{Conserved sandpile with a variable height restriction}
\author{Vanuildo de Carvalho}
\author{\'Alvaro de Almeida Caparica}\email{caparica@if.ufg.br}
\affiliation{Instituto de F\'{\i}sica, Universidade Federal de
Goi\'as, C.P.131, 74.001-970,
Goi\^ania (GO), Brazil}
\author{Ronald Dickman}
\affiliation{Departamento de F\'{\i}sica
and National Institute of Science and Technology for Complex Systems,
ICEx, Universidade Federal de
Minas Gerais, Caixa Postal 702, 30161-970 Belo Horizonte, MG, Brazil}

\begin{abstract}
We study a restricted-height version of the one-dimensional Oslo sandpile with conserved
density, using periodic boundary conditions. Each site has a limiting height which can be either two
or three. When a site reaches its limiting height it becomes active and may topple, loosing two particles,
which move randomly to nearest-neighbor sites. After a site topples it is randomly assigned a new limiting
height. We study the model using mean-field theory and Monte Carlo simulation, focusing on the
quasi-stationary state, in which the number of active sites fluctuates about a stationary value.
Using finite-size scaling analysis, we determine the critical particle density and associated critical exponents.
\end{abstract}

\keywords{SOC, mean-field theory, computer simulation}

\maketitle

\section{Introduction}
\begin{spacing}{1.5}

Sandpile models are paradigmatic examples of self-organized criticality (SOC) \cite{Bak1987,Dhar1999},
a control mechanism that forces a system with an absorbing-state phase transition to its critical
point \cite{Dickman2000,Munoz2002,Dickman2002}, without explicit tuning of control
parameters \cite{Grinstein1995}. SOC in a slowly driven sandpile corresponds to an absorbing-state phase
transition in a model with the same local dynamics, but a fixed number of
particles\cite{Dickman2000, Tang1988, Vespignani1997, Vespignani1998, Dickman1998,AVespignani1998}, so-called
{\it conserved sandpiles} \cite{Dickman1998, Munoz1999, Chessa1998, Montakhab1998}.
Absorbing-state phase transitions arise in the context of spatial stochastic models,
and correspond to a transition between an active, fluctuating phase, and an absorbing one, which
allows no escape \cite{Harris1974,Grassberger1979,marro}.

Sandpile models with probabilistic toppling rules, typified by the Manna model \cite{Manna1990,Manna1991},
are commonly designated as {\it stochastic sandpiles}; their study has been central to establishing the
connection between SOC and absorbing-state phase transitions. An important stochastic model is the Oslo
model \cite{Christensen2002}, inspired by experimental studies on rice piles.  In this work we study a
conserved version of the Oslo model, characterizing its absorbing-state critical point.

An inconvenient feature of many sandpile models is the absence of an upper bound on the number of particles
that may occupy a given site, which complicates theoretical approaches such as $n$-site approximations or
continuum descriptions.  This motivated the study of {\it restricted} sandpiles \cite{RDickman2002}.
In the present work we impose a height
restriction on the conserved Oslo model.
Since the symmetries and conserved quantities of the restricted and unrestricted models are the same,
one expects, on the basis of experience with critical phenomena both in and out
of equilibrium, that
the models belong to the same universality class, as is indeed borne out for conserved
versions of the Manna model \cite{munoz08,mnrst2,sharon}.
The symmetries here are limited to spatial translation and inversion, while the
conservation law is that of particle number.
This universality class has come to be known as
the conserved directed percolation (CDP) class. On this basis, it would be most surprising if the restricted Oslo model were to belong to a different universality class that its unrestricted counterpart.  This question
nevertheless merits investigation via numerical simulation. Recently it was suggested that the critical behavior
of conserved stochastic sandpiles in fact belongs to the (non-conserved) directed percolation class \cite{basu},
but further studies are required to verify this assertion.

The remainder of this paper is organized as follows:
In Section II we describe the model and in Section III develop a one-site mean-field approximation.
Our numerical results are reported in Section IV and in Section V we present our conclusions.

\section{Model}

We study a restricted sandpile model with a variable height limit, defined on a lattice of $N_{site}=L^{d}$ sites,
where $d$ is the dimension of space. The configuration is specified by the particle numbers
$z_i$ ($i = 1,\ldots,N_{site}$) at each site. Each site has a {\it critical height} $c_i \in \lbrace 2, 3\rbrace$,
with equal probability, such that $z_i \leq c_i$. A site with $z_i = c_i$ is said to be {\it active}.
Any configuration devoid of active sites is {\it absorbing}, i.e., it admits no escape.
The dynamics of the model proceeds via {\it toppling} of active sites; each active site
has a rate of unity to topple.  (By a "rate of unity" we mean that the time unit is chosen such that
if there are currently $N_A$ active sites, then the time increment associated with the
next toppling is $1/N_A$.)
In a continuous-time (sequential) dynamics, each active site has the same
probability of being the next to topple. When a site, say $i$, topples, two particles are transferred from
$i$ to sites $j$ and $j'$, nearest neighbors of
$i$. The two sites are chosen at random, independently, from the set of nearest neighbors,
and so are not necessarily distinct; we refer to this procedure as an {\it independent} toppling rule.
Due to the height restriction, any particle transfer that would result in a target site $j$
having $z_j > c_j$ is rejected.
(This means that the configuration $z_i = c_i$, $\forall i$ is also absorbing, since no particles can be transferred. The particle densities of interest in this study, however, remain far below the density
associated with this configuration.) When site $i$ loses a particle or particles due to toppling, a new limiting height $c_i$ is selected, equal to 2 or 3, each with probability 1/2. Thus the dynamics has three stochastic elements: (1) the choice of the next site to topple; (2) the choice of target sites $j$ and $j'$ for particle transfers; (3) the choice of the new limiting height after a site topples.

In practice the next site to topple is selected at random from a list of currently active sites,
which must naturally be updated following each toppling event. The time increment associated with
each toppling (whether particles are transferred or not) is $\Delta t = 1/N_a$, with $N_a$ the number
of active sites immediately prior to the event.

\section{Mean-field theory}

The primary aim of the mean-field analysis is to obtain a preliminary idea of the phase diagram and
(assuming the latter possesses a phase transition), an order of magnitude estimate of the critical
point.
We consider the simplest mean-field approach, known as the one-site approximation. At this level of
approximation, there are seven possible states $(i|j)$ for a given site, where $i=0, 1,...,j$ represents the occupation number
and $j=2, 3$ denotes the limiting height, with associated probabilities
denoted by $P_{ij}$.  Taking into account the conditions of normalization
\begin{equation}\label{a1}
\sum_{j=2}^{3}\sum_{i=0}^{j}P_{ij} = 1,
\end{equation}
and of fixed density,
\begin{equation}\label{a2}
\sum_{j=2}^{3}\sum_{i=0}^{j}iP_{ij}=\zeta,
\end{equation}
there are only five independent variables at this level.

\begin{figure}[!htb]
\begin{center}
 \includegraphics[width=0.60 \linewidth]{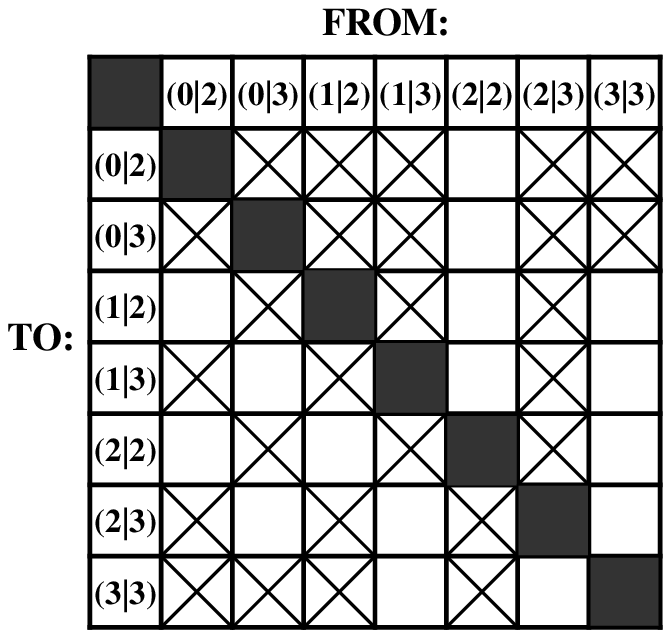}
\caption{Transitions between states of a single site. Transitions marked ``X'' are impossible and those ones related to
diagonal elements are irrelevant.}\label{Transitions}
\end{center}
\end{figure}

We begin the analysis by listing the possible transitions between states $(i|j)$ in
Fig.~\ref{Transitions}. Each transition (at a given site, called the {\it central site} in
this discussion), requires a specific configuration at the central site and at one or both of its nearest
neighbors, and a certain redistribution of particles from the toppling site.  (The local configuration and
the choice of target sites, $j$ and $j'$, in the particle redistribution are statistically independent events.)
In the one-site approximation, joint probabilities involving two or more sites
are factorized.  Denoting a joint two-site probability by $P_{ij|kl}$, the one-site approximation uses
the replacement $P_{ij|kl}\rightarrow P_{ij}P_{kl}$, and similarly for three-site
probabilities.

To illustrate how the rates associated with these transitions are calculated, we discuss some examples. Consider first the transition $(0|2)\rightarrow (1|2)$.  The initial configuration must be either $(02|22)$ or $(03|23)$, that is, the central site must be vacant, have $z_c=2$, and have an active neighbor. When the latter topples, exactly one particle must migrate to the central site.  On a hypercubic lattice in $d$ dimensions, each site has $2d$ nearest neighbors.
Since the probability of exactly one particle jumping to the central site is $2 (1/2d) [1- (1/2d)]$,
the rate (per site) of transitions of the kind  $(0|2)\rightarrow (1|2)$ is
\begin{equation}
2d\frac{2d-1}{2d^{2}}P_{02}(P_{22}+P_{33}),
\end{equation}
where the factor $2d$ represents the number of nearest neighbors.

Consider next the transition $(2|2)\rightarrow (0|2)$, which can occur via two mutually
exclusive paths.  In one, both particles liberated when the central site topples attempt to migrate to
the same neighbor, an event having probability $1/4d^{2}$.  In order for both particles to actually migrate, the difference $z_{c}-z$ at the target site must be greater than one.  Thus the initial configurations for which this transition may occur
are $(00|23)$, $(02|32)$, and $(12|32)$. The transition rate for this path is
\begin{equation}
 \frac{1}{2}2d\frac{1}{4d^{2}}P_{22}(P_{02}+P_{03}+P_{13}),
\end{equation}
where the factor $1/2$ represents the probability that the limiting height retains the value of 2
following the toppling event.  In the other path, the two particles migrate to distinct neighbors
of the central site.  The configurations that allow this transition to occur are
$(2^{\dagger}22^{\dagger}|222)$, $(2^{\dagger}23^{\dagger}|223)$,
and $(3^{\dagger}23^{\dagger}|323)$, where $2^{\dagger}$ and $3^{\dagger}$
denote, respectively, sites with $z<2$ and $z<3$. Thus the transition rate for this path is
\begin{equation}
\frac{2d-1}{4d}P_{22}(P_{2^{\dagger}2}+P_{3^{\dagger}3})^{2}.
\end{equation}
Evaluating the rates of the remaining transitions, we find the equations that govern the probabilities
$P_{i,j}$ at this level of approximation. The equations for the $P_{ij}$ are

\begin{eqnarray}
\frac{dP_{02}(t)}{dt}&=&\frac{1}{4d}P_{22}(P_{02}+P_{03}+P_{13}) +\frac{2d-1}{4d}P_{22}(P_{2^{\dagger}2}+P_{3^{\dagger}3})^{2}-\frac{4d-1}{2d}P_{02}\nonumber\\
&&\times(P_{22}+P_{33}),\\
\frac{dP_{03}(t)}{dt}&=&\frac{1}{4d}P_{22}(P_{02}+P_{03}+P_{13}) +\frac{2d-1}{4d}P_{22}(P_{2^{\dagger}2}+P_{3^{\dagger}3})^{2}-\frac{4d-1}{2d}P_{03}\nonumber\\
&&\times(P_{22}+P_{33}),\\
\frac{dP_{23}(t)}{dt}&=&\frac{1}{2d}\left[ P_{03}+2(2d-1)P_{13}-(4d-1)P_{23}\right](P_{22}+P_{33})+\frac{1}{4d}[ 1+2(2d-1)\nonumber\\
&&\times(P_{22}+P_{33})](P_{12}+P_{23})P_{33},\\
\frac{dP_{12}(t)}{dt}&=&\left( \frac{2d-1}{2d}P_{02}-\frac{4d-1}{2d}P_{12}\right)(P_{22}+P_{33}) +\frac{1}{4d}\left[ 1+2(2d-1)(P_{22}+P_{33})\right]\nonumber\\
&&\times P_{22}(P_{12}+P_{23})+\frac{1}{4d}\left[P_{02}+P_{03}+P_{13}+(2d-1)(P_{2^{\dagger}2}+P_{3^{\dagger}3})^{2}\right]P_{33},\nonumber\\
&&\\
\frac{dP_{13}(t)}{dt}&=&\left( \frac{2d-1}{2d}P_{03}-\frac{4d-1}{2d}P_{13}\right)(P_{22}+P_{33}) +\frac{1}{4d}\left[ 1+2(2d-1)(P_{22}+P_{33})\right]\nonumber\\
&&\times P_{22}(P_{12}+P_{23})+\frac{1}{4d}\left[P_{02}+P_{03}+P_{13}+(2d-1)(P_{2^{\dagger}2}+P_{3^{\dagger}3})^{2}\right]P_{33},\nonumber\\
&&
\end{eqnarray}

\begin{eqnarray}
\frac{dP_{22}(t)}{dt}&=&\frac{1}{2d}\left[P_{02}+(4d-1)P_{12}\right](P_{22}+P_{33})+\frac{1}{4d}\left[ 1+2(2d-1)(P_{22}+P_{33})\right]\nonumber\\
&&\times P_{33}(P_{12}+P_{23})-\frac{1}{2d}\left[P_{02}+P_{03}+P_{13}+(2d-1)(P_{2^{\dagger}2}+P_{3^{\dagger}3})^{2}\right]P_{22}\nonumber\\
&&+\frac{1}{2d}\left[ 1+2(2d-1)(P_{22}+P_{33})\right]P_{22}(P_{12}+P_{23}),\\
\frac{dP_{33}(t)}{dt}&=&\frac{1}{2d}\left[P_{13}+(4d-1)P_{23}\right](P_{22}+P_{33})-\frac{1}{2d}\left[ P_{02}+P_{03}+P{13}\right]P_{33}\nonumber\\
&&+\frac{2d-1}{2d}(P_{2^{\dagger}2}+P_{3^{\dagger}3})^{2}P_{33}-\frac{1}{2d}\left[ 1+2(2d-1)(P_{22}+P_{33})\right]P_{33}\nonumber\\
&&\times(P_{12}+P_{23}).
\end{eqnarray}

Solution of the above equations is performed numerically.  We note that in light of the constraints expressed in
Eqs. (\ref{a1}) and (\ref{a2}), we have only five independent equations. We also take advantage of the following
symmetry: in the mean-field approximation, if $P_{02} = P_{03}$ initially, then this equality continues to hold
throughout the evolution.  A similar relation holds between $P_{12}(t)$ and $P_{13}(t)$.
We therefore obtain a set of three independent differential equations for $P_{22}(t)$,
$P_{23}(t)$ e $P_{33}(t)$, which are readily integrated using a fourth-order Runge-Kutta
scheme \cite{Numerical-2007}. We define the order parameter as the fraction of active sites,
\begin{equation}
\rho(t)=P_{22}(t)+P_{33}(t).
\end{equation}

\begin{figure}[!htb]
\begin{center}
 \includegraphics[width=0.8 \linewidth]{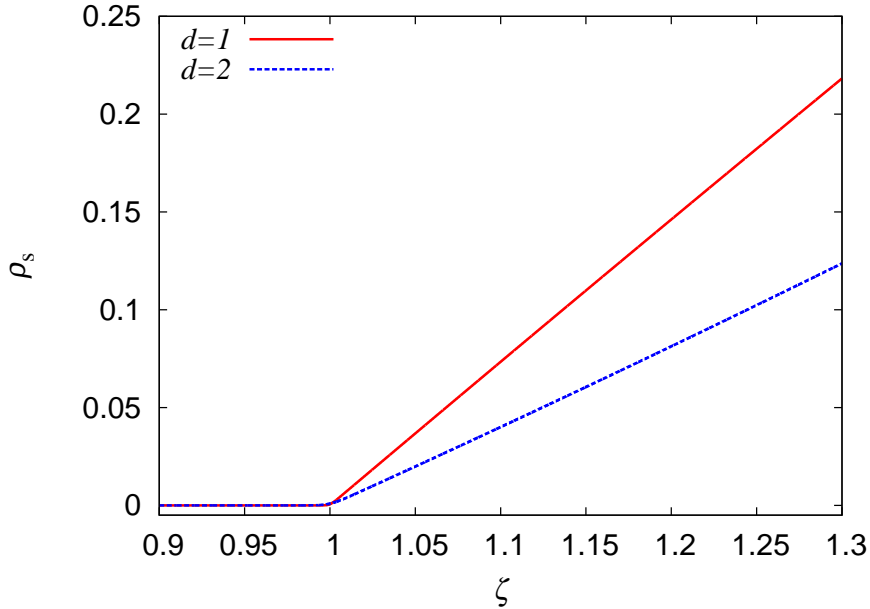}
\caption{Stationary order parameter $\rho_{s}$ versus density $\zeta$,
in the one-site approximation, for $d=1$ and $d=2$.}\label{foto9}
\end{center}
\end{figure}

Numerical integration reveals that in one and two dimensions, $\rho(t) \to 0 $ as $t \to \infty$ for particle
densities $\zeta \leq 1$, while for higher densities it attains a nonzero stationary value $\rho_{s}$
(see Fig.~\ref{foto9}), which grows continuously with $\zeta - 1$. Thus in the one-site approximation, the
model exhibits a continuous phase transition between an active and an absorbing state.  Such a continuous
absorbing-state phase transition is familiar from studies of the contact process \cite{Harris1974}, and
of conserved stochastic sandpiles, among other models.  We verify that for $\zeta \neq 1$, $\rho(t)$
approaches its stationary value exponentially: $|\rho(t)-\rho_{s}| \propto \exp( -t/\tau)$, where the
relaxation time $\tau$ depends on $\zeta$, and diverges as $\zeta \to \zeta_c = 1$, following
$\tau \sim 1/|\zeta -\zeta_c|$, as is typical for mean-field analysis of absorbing-state phase
transitions \cite{marro}. The one-site approximation yields the critical exponent $\beta$, defined via
$\rho_{s}\sim\left(\zeta-\zeta_{c}\right) ^{\beta}$, (for $\zeta > \zeta_c$), as $\beta=1$ for both
$d=1$ and $d=2$.  This value is expected for mean-field analysis of continuous absorbing-state phase transitions
in models that do not possess up-down (or particle-hole) symmetry \cite{marro}.  The reason is that in the
absence of such a symmetry, all powers of the order parameter are allowed in the mean-field equations
of motion, so that near the critical point, the terms proportional to $\rho$ and $\rho^2$ dominate
(that is, $d\rho/dt \simeq a\rho - b\rho^2$, with $a \propto \zeta-\zeta_c$ and $b>0$), and the stationary value
of $\rho$ is proportional to $\zeta-\zeta_c$.

\section{Simulation}

We simulate the restricted sandpile model described above in one dimension using periodic boundaries, on rings
of $L=500$, 1000, 1500, and 2000 sites. The initial configuration is defined by assigning limiting heights
$z_c =2$ or 3 with equal probabilities, independently, to each site, and then distributing randomly $N$
particles among the $L$ sites, avoiding occupancies that exceed the maximum height.  The resulting
{\it initial} distribution
is statistically homogeneous; the occupations of different sites are essentially independent. Once all $N$
particles have been inserted, the stochastic dynamics, which, as noted conserves particles, begins. For each
system size $L$, we study an interval of densities $\zeta=N/L$. In all cases, we use $N_{r}=2000$ independent
realizations of the process. The maximum time is $t_{m}=5\times10^{4}$ units for $L=1000$ and
$t_{m}=3\times10^{5}$ for $L=2000$.

\begin{figure}[!htb]
\begin{center}
 \includegraphics[width=0.8\linewidth]{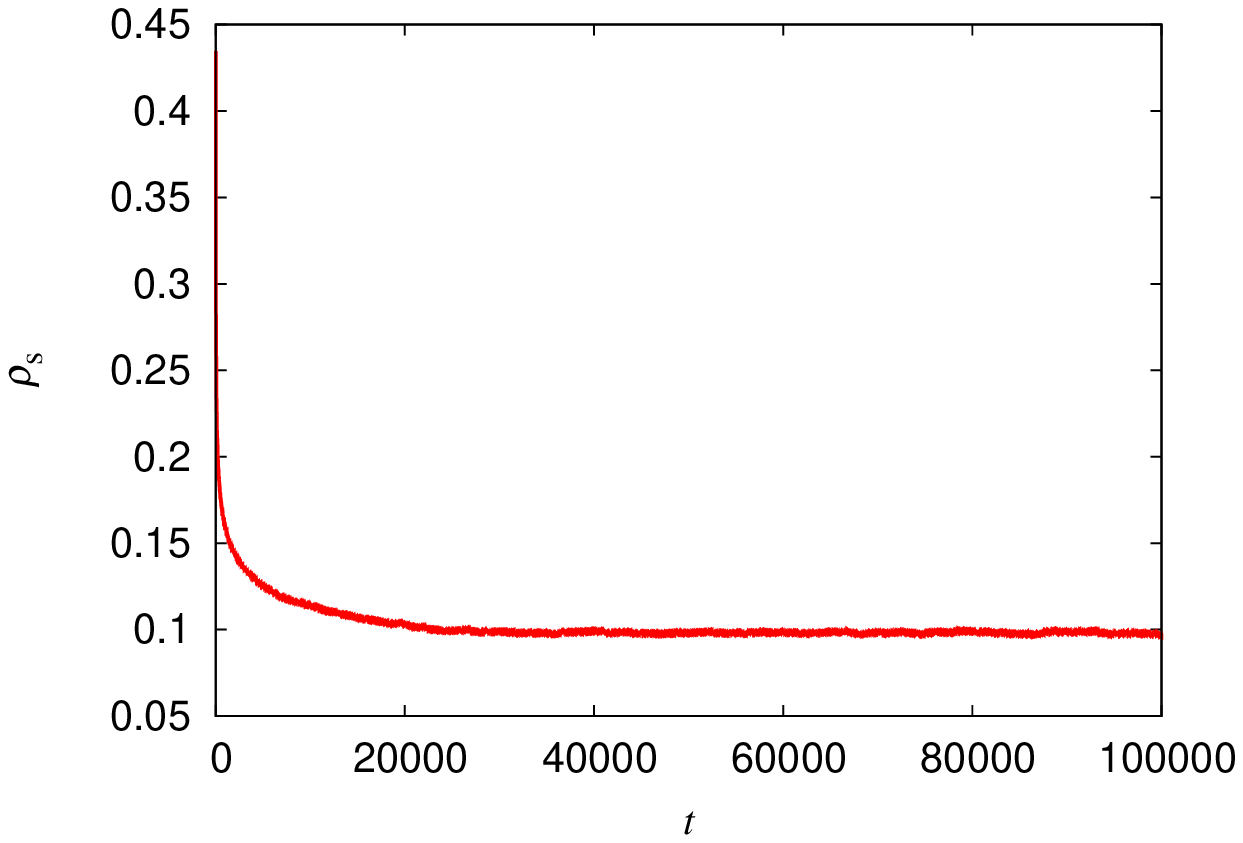}
 \includegraphics[width=0.8\linewidth]{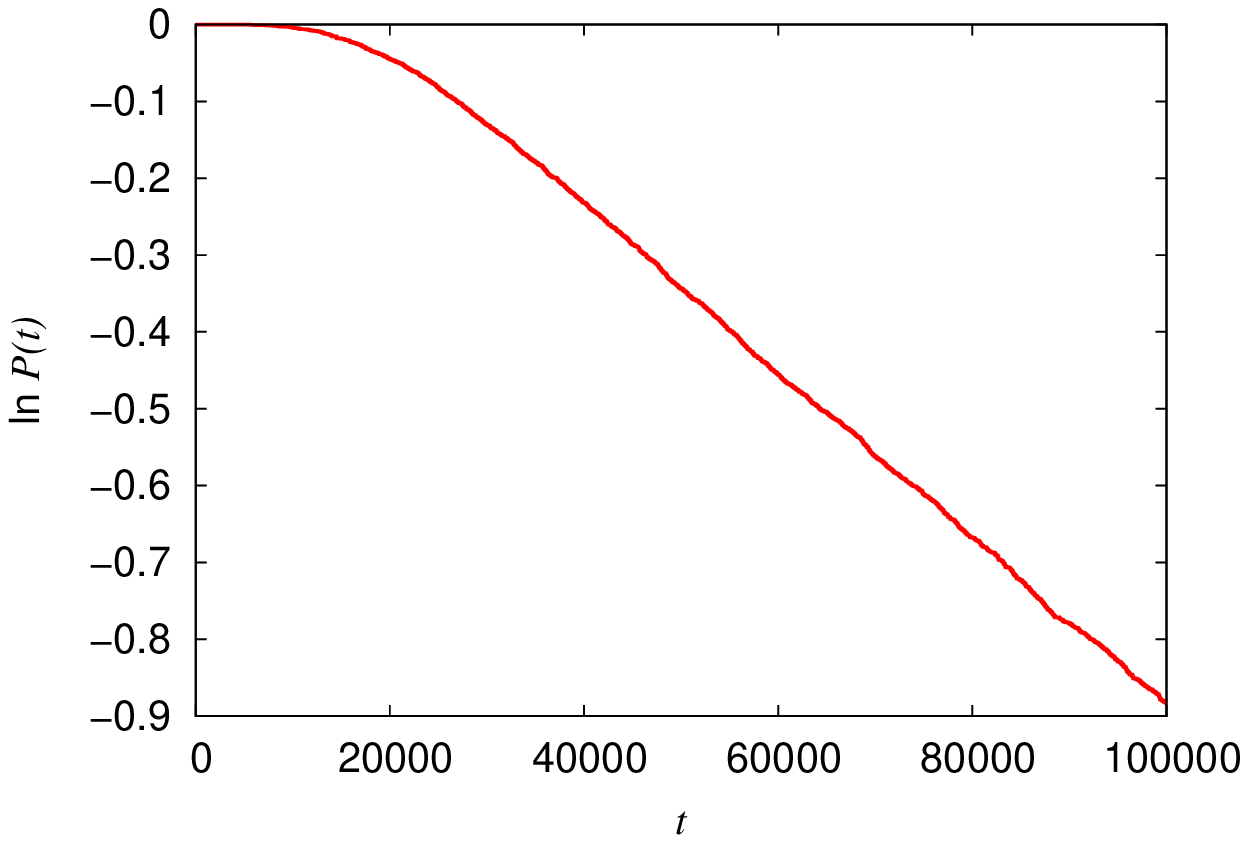}
\end{center}
\caption{Simulation: order parameter (upper panel) and survival probability (lower panel) versus time for $L=1000$ and particle density $\zeta=1.64$.}\label{foto4}
\end{figure}

To determine the critical behavior of the one-dimensional version of sandpile defined above, we study the time-dependent density of active sites $\rho(t)$ as well as their survival probability $P(t)$.  Figure ~\ref{foto4} shows the typical simulation behavior for $\rho(t)$ and $P(t)$. We see that $\rho(t)$ possesses a transient part before reach a well defined stationary value $\rho_{s}(\zeta,L)$, while the survival probability $P(t)\propto \exp(-t/\tau(\zeta,L))$ has an exponential decay.  Discarding the
initial transient portion of the data the survival time $\tau(\zeta,L)$ is estimated by the slope of the curve.

\begin{figure}[!htb]
\begin{center}
 \includegraphics[width=0.8 \linewidth]{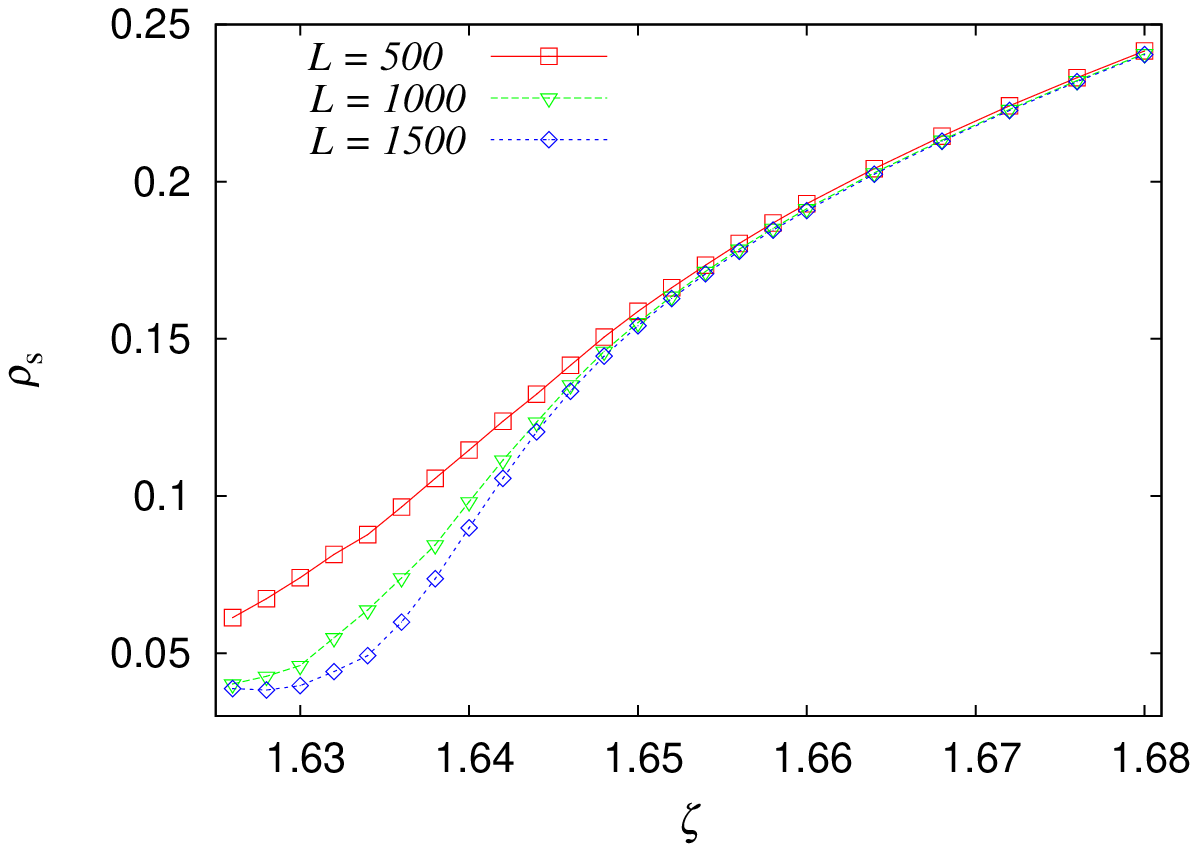}
\caption{Simulation: order parameter $\rho_{s}$ versus particle density $\zeta$ for sizes as indicated.}
\label{foto2}
\end{center}
\end{figure}

In simulations, the particle density $\zeta$ cannot be varied continuously; for each system size $L$ it can only be changed in increments of $1/L$. To have access to intervals of particle density smaller than $1/L$, we follow a method employed in the study of conserved sandpile models \cite{RDickman2002} and pair contact process \cite{Rabelo2002}. Initially we determine the stationary average of $\rho$ for a series of discrete values of the particle density, as shown in Fig.~\ref{foto2}. Since it is reasonable to suppose that the resulting points fall on a smooth curve (as is indeed confirmed by the data), we then use quadratic interpolation to estimate $\rho$ at particle densities that are not accessible for the sizes studied here.

\begin{figure}[!htb]
\begin{center}
 \includegraphics[width=0.75 \linewidth]{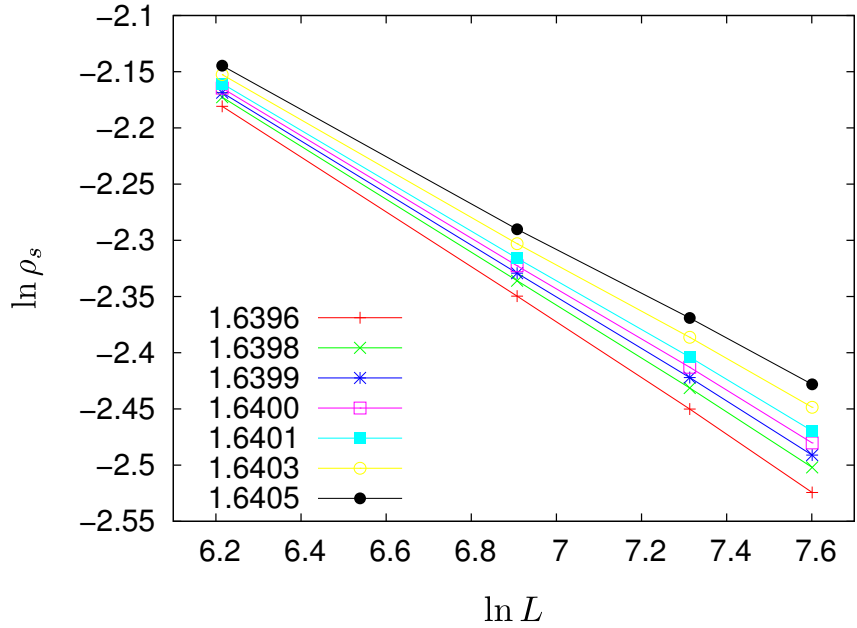}
\caption{Stationary order parameter $\rho_{s}$ versus system size $L$ for particle densities $\zeta$ as
indicated.}\label{foto3}
\end{center}
\end{figure}

By means the analysis of these data, we obtain the order parameter and the mean survival time as functions of system size for diverse values of the particle density, as shown in Fig.~\ref{foto3}. At an absorbing-state phase transition, the critical point of a phase transition is determined by seeking a power-law dependence of the order parameter $\rho_{s}$ and the survival time $\tau$ on the system size $L$. These two parameters are governed by
\begin{eqnarray}
 \rho_{s}(\zeta,L)&=&L^{-\beta/\nu_{\bot}}\textbf{C}(L^{1/\nu_{\bot}}\Delta),\\
 \tau(\zeta,L)&=&L^{\nu_{\parallel}/\nu_{\bot}}\textbf{R}(L^{1/\nu_{\bot}}\Delta),
\end{eqnarray}
where $\Delta\equiv\zeta-\zeta_{c}$ is the distance from criticality and \textbf{C} and \textbf{R} are finite-size scaling relations \cite{Grassberger1979}. At the critical point ($\Delta=0$), we expect $\rho_{s}(\zeta_{c},L) \sim L^{-\beta/\nu_{\bot}}$ and $\tau(\zeta_{c},L) \sim L^{\nu_{\parallel}/\nu_{\bot}}$.
With this in mind, we can estimate $\zeta_{c}$ and $\beta/\nu_{\bot}$ from the curve of $\rho_{s}(\zeta_{c},L)$ that best approximates a straight line when plotted versus $L$ on log scales.

\begin{figure}[!htb]
\begin{center}
 \includegraphics[width=0.8\linewidth]{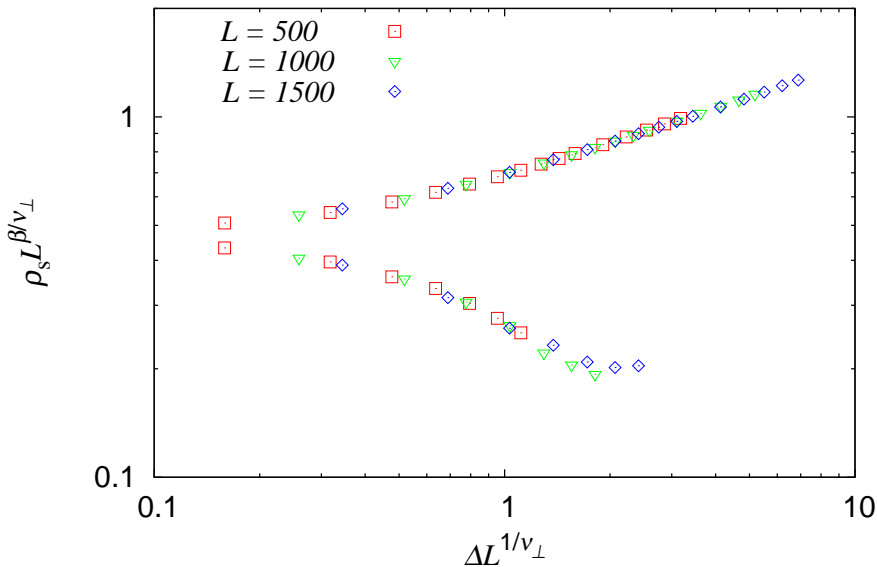}
\caption{Scaling plot for the density of active sites.}\label{foto1}
\end{center}
\end{figure}

This analysis yields $\zeta_{c}=1.6400(2)$ and $\beta/\nu_{\bot}=0.227(5)$, where the figures in parentheses
denote the uncertainty in the last significant figure. Analyzing the data for the lifetime $\tau$ in the same
manner, we obtain $z=\nu_{\parallel}/\nu_{\bot}=1.44(3)$. (The uncertainties are related to two contributions: one due to the uncertainty of the fit, the other due to the uncertainty in the values of $\rho_{s}$ and $\tau$ for each size $L$.) We estimate the critical exponent $\nu_{\bot}$ by
plotting $L^{\beta/\nu_{\bot}}\rho_{s}(\zeta,L)$ versus $L^{1/\nu_{\bot}}\Delta$ for various system sizes,
seeking the value of $\nu_{\bot}$ which yields the best data collapse. Figure~\ref{foto1} shows that a good
collapse is obtained using $1/\nu_{\bot}=0.704(5)$. The critical exponent $\beta$ that relates the order parameter $\rho_{s}$ with $\Delta$ through the relation $\rho_{s}=\Delta^{\beta}$ is then easily determined as $\beta=0.322(5)$.

Table~\ref{Exponents} compares our estimates for critical exponents with those obtained in studies of other
one-dimensional models in the CDP universality class.  Despite apparent differences, it is important to recall
that previous studies have revealed that simulations of large systems (20$\,$000 sites or larger) are needed
to obtain reliable values of critical exponents for this class \cite{rstmanna06,srw}. For example,
studies using smaller system sizes overestimated the value of the critical exponent $\beta$
in the conserved Manna sandpile, in both its restricted and unrestricted versions \cite{Dickman2000,RDickman2002}.

\begin{table}
\centering
\caption{{\footnotesize Critical exponents for one-dimensional models in the CDP universality class compared
with estimates from present work. $^{\rm a}$ Restricted Manna model \cite{rstmanna06};
$^{\rm b}$ CDP Field theory \cite{cdpft};
$^{\rm c}$ Sleepy Random Walkers \cite{srw}.}}\label{Exponents}
\begin{tabular}{llllll}
\hline
\hline
Model            & $\beta/\nu_\bot$ &   $z$   & $\beta$ \\ \hline
Rest. Manna$^{\rm a}$  &     0.213(6)     & 1.55(3) & 0.29(1) \\
CDP - FT$^{\rm b}$     &     0.214(8)     & 1.47(4) & 0.28(2) \\
SRW$^{\rm c}$          &     0.212(6)     & 1.50(4) & 0.290(4)\\
Present work           &     0.227(5)      & 1.44(3) & 0.322(5) \\
\hline
\hline
\end{tabular}
\end{table}

\section{Conclusions}

We study a height-restricted fixed-density version of the Oslo sandpile in one dimension. At each site, the
limiting height $z_{c}$ may be either 2 or 3. The model is found to exhibit a continuous phase transition
between and active and an absorbing state at a critical value of the particle density, $\zeta_{c}$. The
one-site mean-field approximation predicts $\zeta_c = 1$, whereas simulations yield
$\zeta_{c}=1.6400(2)$.  The small sizes analyzed here limit the reliability of our estimates for the
critical exponents. Comparison with literature values (Table I) raises the
possibility that the restricted Oslo model does not belong to the conserved directed percolation class.
More definitive conclusions will however require studies of larger systems.

\bibliographystyle{unsrt}
% \bibliography{referencias}

\end{spacing}

\end{document}